\documentclass[12pt]{article}

\usepackage{epsfig}
\usepackage{graphicx}
\usepackage{amssymb,amsmath}

\usepackage{epsfig,cite}
\usepackage{color}
\usepackage{graphics}
\usepackage{amsfonts}
\usepackage{epsfig}
\def\one{{\hbox{1\kern-.8mm l}}}

\def\td{\tilde }

\def\non{\nonumber}

\newcommand{\beq}{\begin{equation}}
\newcommand{\eeq}{\end{equation}}
\newcommand{\be}{\begin{eqnarray}}
 \newcommand{\ee}{\end{eqnarray}}

\newcommand{\ov } {\over }
\newcommand{\p }{\partial }

\newcommand{\s }{\sigma }

\def\td{\tilde }
\def\a{\alpha }
\def\k{\kappa }
\def\dk{\tilde\kappa }
\def\xe{\varepsilon}
\def\txe{\tilde\varepsilon}
\def\ta{\tau }

\def\appendix#1{
  \addtocounter{section}{1}
  \setcounter{equation}{0}
  \renewcommand{\thesection}{\Alph{section}}
  \section*{Appendix \thesection\protect\indent \parbox[t]{11.15cm}
  {#1} }
  \addcontentsline{toc}{section}{Appendix \thesection\ \ \ #1}
  }

\begin{document}
\null\vskip-24pt 
\hfill
UB-ECM-PF-04/30
\vskip-1pt
\hfill
SISSA-72/2004/EP 
\vskip-1pt
\hfill {\tt hep-th/0410152}
\vskip0.2truecm
\begin{center}
\vskip 0.2truecm {\Large\bf
Search for the most stable massive state\\
\vskip 0.2truecm
 in superstring theory}
\\
\vskip 2.5truecm

{\bf Diego Chialva $^{a,b}$
, Roberto Iengo $^{a,b}$ 
and Jorge G. Russo $^{c}$
}

\bigskip
\medskip


{{${}^a$
\it International School for Advanced Studies (SISSA)\\
Via Beirut 2-4, I-34013 Trieste, Italy} 

\medskip

{${}^b$ INFN, Sezione di Trieste}


\medskip

{${}^c$  
Instituci\' o Catalana de Recerca i Estudis Avan\c{c}ats (ICREA),\\
Departament ECM,
Facultat de F\'\i sica, Universitat de Barcelona,  Spain}
}
\bigskip

{\tt{chialva@sissa.it},
{iengo@he.sissa.it},
{jrusso@ecm.ub.es}
}

\end{center}
\begin{abstract}

In ten dimensional type II superstring, all perturbative 
massive states are unstable, typically with a  short 
lifetime compared to the string scale.
We find that the lifetime of the average string state of  mass $M$
has the asymptotic form $\bar {\cal T}\leq {\rm const.}g_s^{-2}\ M^{-1}$.
 The most stable string state seems to be a 
certain state with high angular momentum
which can be classically viewed as a circular string rotating in several planes
(``the rotating ring''), predominantly decaying by radiating soft 
massless NS-NS particles,
with  a 
lifetime ${\cal T}= 
c_0 g_s^{-2 } M^ 5$.
Remarkably, the dominant channel is the decay into a similar rotating
ring state of smaller mass. The total lifetime to shrink to zero size is $\sim M^7$.
In the presence of D branes, decay channels involving 
open strings in the final state
are exponentially suppressed, so the  
lifetime is still proportional to $M^5$, except for a D brane at a special 
 angle or flux. 
For large mass, the spectrum for massless emission 
exhibits qualitative features typical of a thermal spectrum, such as
a maximum and an exponential tail.
We also discuss the decay properties of rotating rings
 in the case of compact dimensions.

\end{abstract}

\date{October 2003}

\vfill\eject

\tableofcontents


\setcounter{section}{0}


\section{Introduction}
\setcounter{equation}{0}


Understanding the decay properties of massive string states
may uncover new aspects of string theory dynamics, and 
it may also have direct implications in cosmology 
(see e.g. \cite{witten,myers,Damour,kibble}).
According to their decay properties, the quantum states of the type II string spectrum
can be divided into two classes:

\smallskip

\noindent {\cal i}) Those for which the dominant decay channel is by massless emission.

\smallskip

\noindent {\cal ii})  Those for which the dominant decay channel is by  emission 
of massive particles.

\smallskip

In  \cite{CIR} we argued that massless emission is the dominant channel whenever the quantum state 
corresponds to large closed strings which cannot break during the evolution.
An example of this is the rotating ring: the string is circular, and there is never a contact between two points of the string.
For these class {\cal i)}  string states, the decay channels into massive states
are exponentially suppressed, we indeed find 
$\Gamma_{\rm massive}=O(e^{-c M^ 2})$,
where the value of $c$ depends on the specific state.\footnote{ 
The numerical constant $c>0$ can be  
very low if there are two points of the string that are close
to each other. In such a case the decay into massive states can be viewed semiclassically 
as tunnelling effect.}
An example of a class {\cal ii)} state is the folded closed string corresponding to the quantum state with maximum angular momentum: 
the  points of the string 
are always in contact with the other side so the string can break 
at any moment. This picture was confirmed in detail by the calculations
of \cite{CIR}. With this line of argument, in  \cite{CIR} 
we have identified a one-parameter family
of  long-lived string states.

For the class {\cal i)} quantum states, the lifetime is 
given by ${\cal T^{\rm (i)}}=\Gamma_{\rm massless}^{-1}$, while
for the class {\cal ii)} quantum states, the decay rate for massless emission $\Gamma_{\rm massless}$ 
puts an upper bound on the lifetime, i.e. $
{\cal T^{\rm (ii)}}\leq \Gamma_{\rm massless}^{-1}
$.
In this paper we will concentrate on the calculation of $\Gamma_{\rm massless}$,
first finding a simple general formula, 
and then carrying out the analytic calculation in
detail for the case of the rotating ring.
The analytic calculation confirms the lifetime ${\cal T}_{\rm ring}= 
{c_0\over g_s^2 } M^ 5$, in ten uncompact spacetime dimensions, obtained in  \cite{CIR}  by numerical evaluation of
the imaginary part of the one-loop correction to the two-point function
(here we also determine $c_0$). Further, we show that for large $M$ the decay occurs by the emission
of massless quanta of feeble energy $\omega\sim 1/M$ and that  
the ring final state is just the same
as the initial one, only slightly shrunk to conserve energy. Moreover, we show that for large mass a classical computation gives exactly the same
results as the quantum one. That is, the rotating ring slowly radiates as a classical antenna (section 3.3). 
In section 4 we show that  the rotating ring has the same 
lifetime ${\cal T}_{\rm ring}= 
{c_0\over g_s^2 } M^ 5$ in the presence of generic configurations of D branes. 
This is due to the fact that the decay into
 open strings is exponentially suppressed, except for a D brane at a special angle.

The case of compact dimensions is discussed in section 5. 
The decay properties of the ring then depend on the ring 
configuration, in particular,
on whether the ring lies on compact or uncompact dimensions.
For configurations where decay into winding modes is suppressed, 
the lifetime of the rotating ring is 
${\cal T}= {\rm const.} g_s^{-2}\ M^ {d-4}$, 
where $d$ is the number of uncompact spatial dimensions, while the total lifetime for the ring to shrink to microscopic size is
  ${\cal T}_{\rm tot}= {\rm const.} g_s^{-2}\ M^ {d-2}$.

In section 6, we   compute  $\bar \Gamma_{\rm massless}$ representing the average over all initial states
of given mass $M$ (following the similar calculation carried out in \cite{Amati2} for the bosonic string).

Some earlier discussions of decays of massive string states
can be found in  \cite{Dai,Mitchell,Okada}.
The average decay rate by massless emission
in bosonic string theory 
is given in \cite{Amati2} and a study of massive emission from an average string state is in \cite{Manes}
(averages involving two massless vertices were discussed in \cite{Manes2}).
The classical breaking of the string and the correspondence
with the quantum calculation is discussed in  \cite{IR2}.
A numerical study of the decay of the rotating ring in compactified dimensions
from the one-loop two-point function is in \cite{CI}.

\section{General formula for decay rates}

\setcounter{equation}{0}

The total decay amplitude of a state $|\Phi_0\rangle $ of mass $M$
for emission of a massless (fermion or boson) particle $(\omega,\vec k)$
represented by a vertex operator $V$ is given by
\beq
\Gamma (\omega )={g_s^ 2\over 8M} \int {d\Omega^{d-1} d\omega \ \omega^{d-2}\over 
(2\pi)^{d-1} E'} \delta(M-E'-\omega)\ \sigma_R\ \sigma_L\ 
={g_s^ 2\over 8(2\pi)^{d-1}}{\omega^{d-2}\over M^2}\int d\Omega^{d-1}\ \sigma_R\ \sigma_L\ ,
\label{dcy}
\eeq
where $d$ is the number of spatial dimensions  and
\be
\sigma_R= 
\sum_{\Phi_{N'}}\langle \Phi_0|
V_R(k,1)^\dagger |\Phi_{N'} \rangle \langle\Phi_{N'}|V_R(k,1)
| \Phi_  0 \rangle\ , 
\ee
and a similar expression for $\sigma _L$.
The sum goes over all states with mass $M'=\sqrt{N'}$, with energy
 $E'=\sqrt{{M'}^2+\omega^2}$, and (in  units $\alpha '=4$) 
 $$
\omega={M^2-{M'}^ 2\over 2M}={N_0\over 2\sqrt{N}}\ ,\ \ \  M^ 2=N\ ,\ \ 
N_0 \equiv N-N'\ .
$$
Introducing  a projector onto states of level $N'$, 
\be \label{Nlevelprojector}
 \sum_{\Phi_{N'}} |\Phi_{N'} \rangle \langle\Phi_{N'}|=\oint {dz\over z}\
 z^{\hat N_R-{N'}}\ , 
\ee
the decay amplitude can be computed as
\be
\sigma_R= \oint {dz\over z}\ z^{-{N'}} 
\langle\Phi_0| V_R(k,1)^\dagger  z^{\hat N_R} V_R(k,1) |\Phi_0  \rangle\ .
\ee
By using the standard commutators
$$
[a,a^\dagger]=1\ ,\ \ \ \{ \psi,\psi^\dagger\} =1\ ,\ \ \  
z^{na^\dagger a} e^ {a}=  e^ { a/z^n} z^{na^\dagger a}\ ,\ \ \  
z^{r\psi^\dagger \psi} e^ { \psi }=  e^ { \psi/z^r} z^{r\psi^\dagger \psi}\ ,
$$
one can write 
\be
z^{\hat N_R} V_R(k,1)  = V_R(k,z)z^{\hat N_R}\ .
\ee
Thus
\be
\sigma_R= \oint {dz\over z}\ z^{N-{N'}} 
\langle\Phi_0| V_R(k,1)^\dagger   V_R(k,z) |\Phi_0  \rangle\ . 
\label{sgr}
\ee
For massless NS-NS emission, one considers the vertex operator
$V=g_s\ V_L V_R$, with
\be \label{masslessspin2}
V_R=(\xi.\p X_R + i\ \xi.\psi_R \ k.\psi_R) \ e^ {ik.X_R} ,  \ \ 
V_L=(\tilde \xi .\bar \p X_L + i\ \tilde \xi .\psi_L \ k.\psi_L) \ 
e^ {ik.X_L} , 
\label{mvert}
\ee
\be
\xi.k=\tilde \xi .k=0\ ,\ \ \ \ k^ 2=0\ .
\label{pola}
\ee
We will work in the gauge where $\xi_0=\tilde\xi_0=0$.
Equations (\ref{dcy}), (\ref{sgr}), allow to compute the decay rate of any
quantum state of the string spectrum
by computing  a matrix element involving standard
free/creation annihilation operators.
In the next section we carry out this calculation for the rotating
ring. The calculation can also be carried 
out for other states in a straightforward way.

.

\section{ The rotating ring}
\setcounter{equation}{0}
\subsection{ Decay rate} 

We now consider the following state
\beq
|\Phi_0\rangle= {{b^\dagger}^N\over\sqrt{N!}}\psi_{-1/2}^ {Z_1} 
| 0 \rangle _R  \ \times\  
{ ({\tilde c} ^\dagger)^N\over\sqrt{N!}}\tilde \psi_{-1/2}^ {Z_2}  | 0 \rangle _L\ ,
\eeq
where we  define $b^\dagger \equiv b_{1+}^\dagger ={\alpha_{-1}^{X_1} + i\alpha_{-1}^{X_2}\ov \sqrt{2}}$, 
 $\tilde c^\dagger \equiv \tilde c_{1+}^\dagger={\tilde \alpha_{-1}^{X_3} + i\tilde \alpha_{-1}^{X_4}\ov \sqrt{2}}$
(see appendix A for notation).
The operator $b^\dagger $  ($\tilde c^\dagger $)
adds one unit of Right (Left) angular momentum in the plane 12 (34).
This state has
\beq
N_R=N_L=N\ ,\ \ \ \ \a ' M^2=4N\ ,\ \ \ 
\eeq
\beq
J_{12R}=N , \ \ \  J_{12L}=0  , \ \ \ \ 
J_{34R}=0 , \ \ \  J_{34L}=N  , 
\eeq
so that $J_{12}=J_{34}=N$. 
The classical string soliton solution 
with the same values of $J_{12R},J_{12L}$, $J_{34R},
J_{34L}$ is as follows 
\beq
Z_{1}= L\  e^{-i(\s -\tau )}\ ,\ \ \ \ 
Z_{2}=L\  e^{i(\s +\tau )}\ ,\ \ \ \ \ L=\sqrt{\a' N\over 2}\ ,
\label{cring}
\eeq
$$
X_0= 2\sqrt{2}L\ \tau\ ,\ \ \ \ \ \sigma\in[0,2\pi)\ .
$$
It describes a circular string rotating 
around its axes in two orthogonal planes. One can also write 
similar rotating ring solutions 
(and the corresponding  quantum states) with rotation in different planes.

The lifetime of this state was studied in \cite{CIR} 
by a numerical evaluation of the imaginary part of
the one-loop two point amplitude. 
It was shown that it goes like ${\cal T}\sim M^ 5$,  and that the dominant
decay channel is massless emission. Other decay channels representing decay into two massive particles  
are suppressed exponentially for large mass of the initial state, i.e. 
the decay rate for these massive channels is $\Gamma\sim \exp (-c M^ 2)$.
Thus in the large mass limit, the only contribution  
which is not exponentially suppressed is decay by emission of
a single massless particle
(multiple massless particle emission is suppressed by extra factors
of $g_s^ 2$).
 This can be computed analytically by using the operator approach,
as explained in section 2.

Let us start considering emission of a NS-NS 
massless particle.
It is convenient to introduce complex notation
\beq
\k= {1\over \sqrt{2}}(k_1+ik_2)\ ,
\ \ \ \ 
\tilde \k= {1\over \sqrt{2}}(k_3+ik_4)\ ,
\label{cpx}
\eeq
\be
\xe = {1\over \sqrt{2}}(\xi_1+i\xi_2)\ ,\ \ \ 
\txe = {1\over \sqrt{2}}(\tilde \xi_3+i\tilde \xi_4)\ .
\label{cpxx}
\ee
Now consider  the computation of $\sigma_R $ (\ref{sgr}),
\be
\sigma_R= \oint {dz\over z}\ z^{N_0} 
\langle 0|{b^N\over\sqrt{N!}}\psi_{1/2}^{\bar Z}
V_R(k,1)^\dagger  V_R(k,z)
{{b^\dagger}^N\over\sqrt{N!}}\psi_{-1/2}^ {Z} | 0 \rangle \ ,
\ee
with $N_0 \equiv N-N'$. 
The product $V_R(1)^\dagger V_R(z)$ gives rise to five  contributions,
\be
\sigma_R=\sigma_1+\sigma_2+\sigma_3+\sigma_4+\sigma_5\ ,
\ee
with
\be
\sigma_1 =-{1\over N!} 
 \oint {dz\over z}\ z^{N_0} 
\langle 0|b^N: e^ {-ik.X(1)} e^{ik.X(z)}\xi.\p X(1)\xi. \p X(z):
{b^\dagger}^N | 0 \rangle \ ,
\label{sss}
\ee
where we have used
$ \langle 0|\psi_{1\over 2}^{\bar Z}
\psi_{-{1\over 2}}^Z | 0 \rangle =1$, and
 $\sigma_2 , ..., \sigma_5$ are given in appendix B.

For large $N$,  $\sigma_1$  turns out to be the dominant contribution,
while  $\sigma_2 , ..., \sigma_5$ are suppressed by inverse powers of $N$
(see Section~3.2).

Using the algebra of creation/annihilation operators, we obtain 
$\sigma_1 =\sigma_1^{(I)}+\sigma_1^{(II)}$, with
\be
\sigma_1^{(I)}=\big( \bar \xe^2 \k ^ 2
+\xe^2\bar   \k ^ 2\big) \oint {dz\over z^ 2}z^{N_0}\sum_{m=0}^ {N-2}
{4\ N!(2\k\bar \k)^ m \over m!(m+2)!(N-m-2)!}{(1-z)^ {2m+2}\over z^ m}\ ,
\non
\ee
\be
\sigma_1^{(II)}=2 \bar \xe  \xe
\oint {dz\over z}z^{N_0}\big( z+z^{-1} \big) 
\sum_{m=0}^ {N-1}
{N! (2\k\bar \k )^ m \over m!^ 2(N-m-1)!}{(1-z)^ {2m}\over z^ m}\ .
\non
\ee

Now we expand the binomial $(1-z)^r$ and perform the integrals in $z$ 
around a small contour near $z=0$. We find
\be
\sigma_1^{(I)}=4(2\k\bar\k )^{N_0-1}\big( \bar \xe^2 \k^ 2
+\xe^2\bar   \k^ 2\big)
\sum_{r=0}^ {N-N_0-1}
c_I(r)\ (-2\k\bar \k )^ r\ ,
\label{saa}
\ee
\be
\sigma_1^{(II)}=4(2\k\bar\k )^{N_0-1} \bar \xe  \xe 
\sum_{r=0}^ {N-N_0}
 c_{II}(r) \ (-2\k\bar \k )^ r\ ,
\label{saaa}
\ee
where
\beq
c_I(r)= {N! (2r+2N_0)! \over 
r!(N-N_0-1-r)!(r+2N_0)!(r+N_0-1)! (r+N_0+1)!} \ ,
\label{cuno}
\eeq
\beq
c_{II}(r)={ N! (2r+2N_0-2)! [r^ 2+(r+N_0)(2N_0-1)]\over 
r!(N-N_0-r)!(r+2N_0)!(r+N_0-1)!^2}
\ .
\label{cdos}
\eeq
Next, we  sum over polarizations, using the  relations
given in appendix A.
We obtain 
\be
\sigma_1^{(I)}=-2 \left( {N_0^ 2v^ 2\over 4N} \right)^{N_0}\ v^ 2
 \sum_{r=0}^ {N-N_0-1}
c_I(r)\ \left(-{N_0^ 2v^ 2\over 4N} \right)^ r \ ,
\ee
\be
\sigma_1^{(II)}=4\left({N_0^ 2v^ 2\over 4N} \right)^{N_0-1} (1-{v^ 2\over 2})  
\sum_{r=0}^ {N-N_0}
  c_{II}(r)\ \ \left(- {N_0^ 2v^ 2\over 4N}   \right)^ r \ .
\ee
where  $v^ 2\equiv\sin^ 2\theta\cos^ 2\alpha\cos^ 2\beta $.
 The Left contributions are similar, changing $\k,\bar \k$ 
by $\dk ,\bar {\dk }$  and 
$\xi^\mu $ by $\tilde \xi^\mu $, with
$2\dk^ 2={N_0^ 2\over 4 N}\ e^{2i\psi}
\sin^ 2\theta\cos^ 2\alpha\sin^ 2\beta $.

Now we consider the product $\sigma_T=\sigma_L\times \sigma_R$ and
perform the integral over the angular part.
Defining
\beq
I(r,s)=\int d^8\Omega (\sin\theta \cos\alpha)^{2r+2s}(\cos\beta)^ {2r}
(\sin\beta )^ {2s}
={(2\pi )^4\sqrt{\pi }\ r! s!\over 8~\Gamma({9\over 2}+r+s) }\ ,
\eeq
we find
\be
\int d^8\Omega \ \sigma_T=\Sigma_{11}+\Sigma_{22}+2\Sigma_{12}\ ,
\ee
with
\be
\Sigma_{11}=4  \left({N_0^ 2\over 4N} \right)^{2N_0}\ 
 \sum_{r,s=0}^ {N-N_0-1} 
 \left(-{N_0^ 2\over 4N} \right)^{ r+s}c_I(r)c_I(s)I(r+N_0+1,s+N_0+1) \ ,
\label{nombre}
\ee
\beq
\Sigma_{22}= 
16   \left({N_0^ 2\over 4N} \right)^{2N_0-2}\ 
\sum_{r,s=0}^ {N-N_0}    \left(- {N_0^ 2\over 4N}   \right)^{ r+s} c_{II}(r)c_{II}(s)J(r+N_0-1,s+N_0-1)\ ,
\eeq
\beq
\Sigma_{12}= -8
   \left({N_0^ 2\over 4N} \right)^{2N_0-1}\ 
\sum_{r=0}^ {N-N_0-1} \sum_{s=0}^ {N-N_0}
   \left( -{N_0^ 2\over 4N}   \right)^{ r+s} c_I(r)c_{II}(s) K( r+N_0-1,s+N_0-1)\ ,
 \eeq
$$
J(m,n)\equiv I(m,n)-{1\over 2} I(m+1,n)-{1\over 2} I(m,n+1)+
{1\over 4}I(m+1,n+1) \ ,
$$
$$
K(m,n)\equiv  I(m+2,n)-{1\over 2} I(m+2,n+1) \ .
$$
Thus the total decay amplitude for emission of a massless NS-NS particle
from the rotating ring is given by
\be
\Gamma (\omega )= {g_s^2\over 8(2\pi)^8 }{\omega^7\over M^2}
\big( \Sigma_{11}+\Sigma_{22}+2\Sigma_{12}\big) \ .
\ee
Using $\omega =N_0/(2M),\, m+\sqrt{N}$, this can be written as
\be
\Gamma (\omega )=g_s^2 \ N^{-{5\over 2}}\ f(N_0,N)\ ,
\ee
with
\be
 f(N_0,N)={g_s^2\over 2^{10}(2\pi)^8 } {N_0^7\over N^2} 
\big(
\Sigma_{11}+\Sigma_{22}+2\Sigma_{12}\big)\ .
\ee
This function  $ f(N_0,N)$ becomes independent
of $N$ for large $N$. This is due to the fact that the main contributions
arise for finite $N_0$ for any $N$ (see sect. 3.4). In fact, it has a maximum at $N_0=3$
irrespective of the value of $N$ (provided $N>9$). For large $N$,
one can then use the Stirling approximation
to write, in $c_I(r), \ c_{II}(r)$, 
\beq
{N!\over (N-m)!}\sim N^m\ .
\label{stirg}
\eeq
Then the $N$ dependence in  $ f(N_0,N)$ fully cancels out.
Indeed, from $c_I(r) c_{I}(s)$ one gets a power $N^{2N_0+r+s+2}$
which cancels all powers of $N$ in 
$\Sigma_{11}$, and similarly for $\Sigma_{12}$ and $\Sigma_{22}$.
Thus we can replace $ f(N_0,N)$ by  $ \bar f(N_0)=f(N_0,\infty)$ so that 
the large $N$ result is
\be
\Gamma (N_0 )=g_s^2 \ M^{-{5}}\ \bar f(N_0)\ .
\label{spw}
\ee
The total decay rate can now be computed by summing over all $N_0$
from 1 to $N$:
\be
\Gamma _{\rm massless}=g_s^2 \ M^{-{5}} \sum_{N_0=1}^N  f(N_0,N)
\cong g_s^2 \ M^{-{5}} \sum_{N_0=1}^\infty \bar  f(N_0)\ ,
\ee
In the next section, we show that massless fermion emission and emission of massless
RR states
are suppressed by factors $1/N$ and $1/N^2$ respectively, so $\Gamma _{\rm massless}$
represents the complete decay rate modulo $1/N$ corrections.
Thus, for large initial mass, the lifetime of the rotating ring is given by
\be
{\cal T}_{\rm ring}=\Gamma ^{-1}_{\rm massless}=c_0g_s^{-2} \ M^{{5}} \ ,\ \ \ \ c_0\equiv
 \sum_{N_0=1}^\infty \bar f(N_0)\cong {8(2\pi)^8\over 12.5}\ .
\ee
This exactly reproduces the result of \cite{CIR} obtained by 
computing the imaginary part of the one-loop two-point function
(modulo the numerical coefficient $c_0$ which had not been determined in  \cite{CIR}).

\subsection{ Dominant decay channel} 


The large $N$ behavior of the various terms giving the massless (radiation) emission  rate
can be understood in brief by the following sample computation. 

\smallskip

Consider  the term  $\langle \Phi_N^{\rm ring} |\hat O\{X \}|\Phi_N^{\rm ring} \rangle $,
 $\hat O\{X\}\equiv  :\xi .\p X(1) \xi.\p X(z) e^{-ik\cdot X(1)+ik\cdot X(z)}: $~,
 appearing in the Right factor.
We can restrict the full operator $\hat O\{X\}$  to 
the part containing $b,b^{\dag}$,
as only this part gives a contribution.
Ignoring the factors $\xi$ and the dependence on $z$, we have
\beq
\langle \Phi_N^{\rm ring} |\hat O\{ X\}|\Phi_N^{\rm ring} \rangle \sim  
\langle 0|{b^{ N}\over\sqrt{N!}} b^{\dagger }e^{-\sqrt{2}\bar  \k b^{\dagger}} 
b\ e^{\sqrt{2} \k b} ~{b^{\dagger N}\over\sqrt{N!}}|0\rangle
= \sum_{m=0}^{N-1}{N!(-2\k\bar \k )^m\over m!^2(N-1-m)!} \ .
\non
\eeq
For large $N$, one can use Stirling formula (\ref{stirg}). One obtains 
\beq
\langle \Phi_N^{\rm ring} |\hat O\{ X \}|\Phi_N^{\rm ring} \rangle\ \to\ N\sum_{m=0}^\infty {(-2\k\bar \k N )^m\over m!^2}= N f_1(N_0 )\ .
\label{efeu}
\eeq
We have used
$2\k\bar \k= k_1^2+k_2^2=\omega^2 v^2={N_0^2v^2\over 4N}$. Thus this matrix element scales like $N$ for large $N$.

Another term (called $\s_5$ in appendix B) comes from the contraction   
\beq
\langle\xi . \p X(1)\xi.\p X(z)\rangle \langle \Phi_N^{\rm ring} |: e^{-ik\cdot X(1)+ik\cdot X(z)}:|\Phi_N^{\rm ring} \rangle \ .
\label{adaa}
\eeq
This gives rise to the following factor:
\beq
\langle 0|{b^{ N}\over\sqrt{N!}}~e^{-\sqrt{2}\bar \k b^{\dagger}}e^{\sqrt{2} \k b} ~{b^{\dagger N}\over\sqrt{N!}}|0\rangle =
\sum_{m=0}^{N}{N!(-2\k\bar \k )^m\over m!^2(N-m)!}\ \to\  f_1(N_0 )\ ,
\label{ada}
\eeq
which is of order 1 at large $N$. Therefore it is subleading by a factor $1/N$
with respect to expression (\ref{efeu}).
The origin of the extra factor of $N$ in (\ref{efeu}) can be traced back to the action of $\p X\p X$ on the ring state $|\Phi_N\rangle $.
In (\ref{adaa}), the operators $\p X\p X$ are contracted and they do not give $N$ dependence.

Similarly, the contribution from the term  in the vertex $(\xi\cdot \psi\ k\cdot\psi)  (\xi\cdot \psi\ k\cdot\psi )$  (computed in appendix B)
will be subleading for a factor $1/N^2$, since there is no $\p X\p X$ operator and there is an extra power $1/N$ coming from $|k|\sim  1/\sqrt{N}$. 

\smallskip

For the same reason,
 decays by massless fermion and RR emission 
  are suppressed.
Indeed, in the picture we are working, the vertex operator for a massless fermion in the RNS sector is given by
$V^f=V_{R}^f V_{L}$, with 
\be
V_{R}^f(u^\alpha ;k^\mu ) =e^{{i\over 2}\varphi_-}\ \bar u^\alpha\Theta_\alpha \  e^ {ik.X_R}\ ,\ 
\ee
$$
\Theta_\alpha = D^1_{\pm{1\over 2}}... D^5_{\pm{1\over 2}}\ ,\ \ \ \ \  
D^m_{\pm{1\over 2}}=e^{\pm {i\over 2}\phi_-^m}\ .
$$
where $\varphi $ and  $\phi^m$ represent, respectively,
 the bosonized $\beta\gamma $ system and the bosonized $\psi^\mu $.
Here $V_L$ is the NS Left vertex operator (see (\ref{mvert})~).

In view of the above discussion, since for these vertices there is no factor $\xi.\p X$, 
emission of $V^f$ is suppressed by a power of order $1/N$ 
with respect to emission of the NSNS massless particle (\ref{mvert}),
and emission of a massless RR state 
$V^f=V_{R}^f V_{L}^f$ is suppressed by a power $1/N^2$.

We have checked this suppression by a numerical evaluation
of massless fermion and RR emission.
The calculation follows  \cite{CIR}, based on evaluating the emission rate from the imaginary part of the 
correlator of two vertices on the torus, and looking at the  contributions from the different spin structures.
We have found that RNS and RR are subleading, by factors $1/N$ and $1/N^2$ respectively, with
respect to NSNS, as expected in view of the above argument.

\smallskip

The previous considerations lead to an important consequence: 
for large initial mass 
the dominant decay channel is that where the 
final state is a rotating ring of lower mass, i.e. similar to the initial state
with the replacement $N\to N-N_0$. 
The leading process is 
\beq
\Phi_N^{\rm ring} \to \Phi_{N-N_0}^{\rm ring} + \gamma (\omega )\ ,
\label{same}
\eeq
where $\gamma$ is the massless particle. 
To see this explicitly, 
we start with the dominant contribution (\ref{sss}). As pointed out above, because of the normal ordering, we can
drop all other mode operators in $X$, keeping only $b$ and $b^\dagger $. 
Defining $Y(z)=X(z)\bigg|_{b,b^\dagger}$ we  write 
\be
\sigma_1 =-{1\over N!} 
 \oint {dz\over z}\ z^{N_0} 
\langle 0|b^N : e^ {-ik.Y(1)}\xi.\p Y(1)   e^{ik.Y(z)}   \xi. \p Y(z):
{b^\dagger}^N | 0 \rangle\ . 
\label{ssss}
\ee
Next, using the above result (see also appendix B) that the contraction of $\p X(1)\p X(z)$ gives rise
to a contribution suppressed by a power $1/N$, we can write 
$$
: e^ {-ik.Y(1)}\xi.\p Y(1)    e^{ik.Y(z)} \xi.\p Y(z):=: e^ {-ik.Y(1)}\xi.\p Y(1):\   : e^{ik.Y(z)}  \xi.\p Y(z):
\left( 1+O\big({1\over N}\big) \right)\ .
$$
Therefore $\sigma_1$ can also be written as
follows:
\be
\sigma_1 =-{1\over N!} 
 \sum_{\Phi_{N'}}
\langle 0|b^N: e^ {-ik.Y(1)} \xi.\p Y(1): |\Phi_{N'}\rangle
\langle\Phi_{N'}| :e^{ik.Y(1)}\xi. \p Y(1):
{b^\dagger}^N | 0 \rangle \ .
\non
\ee
So consider the calculation of 
$
\langle\Phi_{N'}| :e^{ik.Y(1)}\xi. \p Y(1):{b^\dagger}^N | 0 \rangle \ .
$
The operators $e^{ik.Y(1)}$ and   $\p Y(1)$ contain only $b,\ b^\dagger $ oscillators, 
Therefore,
the states $|\Phi_{N'}\rangle$ can only be made by applying $b^\dagger $ to the vacuum. 
But there is a single state with
$\hat N_R|\Phi_{N'}\rangle =N'|\Phi_{N'}\rangle $ made with  $ b^\dagger$, namely 
 $|\Phi_{N'}\rangle =( b^\dagger)^{N'}|0\rangle $.
Thus in the large $N$ limit the leading contribution to the decay rate
in the sum over all  $|\Phi_{N'}\rangle $ comes from
 a circular ring of lower mass $M'=\sqrt{N'}$.

As an independent check of the decay rate of section 3.1, we now compute the matrix element
\beq
A_R=\langle 0 |{b^{N'} \over \sqrt{N'!}}:\xi. \p Y(1) e^{ik.Y(1)}:
{{b^\dagger}^N  \over \sqrt{N!}}| 0 \rangle \ ,
\eeq
corresponding to the leading process (\ref{same}). In the large $N$ limit, this should coincide with
the decay rate of section 3.1 obtained by summing over all final states of mass $M'=\sqrt{N'}$.
We find
\beq
A_R=-2i\sqrt{(N-N_0)!N!} \sum_{s=0}^{N-N_0} { (-\sqrt{2}\k )^{N_0}(-2\k \bar \k)^s\over s!(N-N_0-s)!}
 \bigg[{\bar \xe \k  (N\! -\! N_0\! -\! s)\over (N_0+s+1)!} + 
 {\xe\over 2\k} {1\over (N_0+s-1)!}\bigg] \ .
\eeq
{}For large $N$,  using the Stirling formula (\ref{stirg}) this becomes
\beq
A_R=-2i\sqrt{N}\ 
(-\sqrt{2N}\k )^{N_0} \sum_{s=0}^{\infty} {(-2\k \bar \k N)^s\over s!}
\bigg[{\sqrt{N}\bar \xe  \k  \over (N_0+s+1)!} 
+  { \xe \over  2\sqrt{N} \k}{1\over  (N_0+s-1)!}\bigg]  
\label{domi}
\eeq
Since $\k $ is proportional to $1/\sqrt{N}$, the full expression is proportional to $\sqrt{N}$.
Now consider $|A_R|^2$. This can be written as
\beq
|A_R|^2=\hat \s^{(I)} +  \hat \s^{(II)}\ ,
\eeq
with
\beq
\hat \s^{(I)}=
4N^2 \big(2\k \bar \k N \big)^{N_0-1}(\bar \xe^ 2\k^2 + \xe^ 2\bar \k^2 )
\sum_{r,s=0}^\infty {(-2\k \bar \k N )^{r+s}
\over r! s! (N_0+s+1)!(N_0+r-1)!}
\eeq
\beq
\hat \s^{(II)}=
2N \bar\xe\xe \big(2\k \bar \k N \big)^{N_0\!-\!1}\!
\sum_{r,s=0}^\infty \! {(-2\k \bar \k N )^{r+s}\!\left[ ( 2\k \bar \k N )^2
+(N_0\!+\!s)(N_0\!+\!s\!+\!1)(N_0\!+\!r)(N_0\!+\!r\!+\!1) \right] \over r! s! (N_0+s+1)!(N_0+r+1)!}
\eeq 
By defining a new summation variable $r'=r+s$, and performing the sum over $r$ at fixed $r'$,
we find that $\hat \s^{(I)} ,\   \hat \s^{(II)}$ {\it exactly}
reproduce  $\sigma_1^{(I)}, \ \sigma_1^{(II)}$ given in eqs. (\ref{saa}), (\ref{saaa}), with $c_I(r),\  c_{II}(r)$ 
given by 
the Stirling limit (\ref{stirg}) of (\ref{cuno}), (\ref{cdos}). 
This provides  another derivation of the decay rate of the rotating ring $\Gamma \cong {\rm const} \ M^ 5$,
by computing only the dominant decay channel.

\subsection{Rotating ring as a classical antenna}

Having shown that the dominant decay channel of the rotating ring is 
by massless
emission, it is instructive to see that the exact large mass decay rate can be 
reproduced by regarding the rotating ring as a classical antenna.

The classical radiation from a source $T_{\mu\nu}(x_0,\vec x )$ is given by
\cite{weinberg}
\beq
\Gamma =c{\omega^7\over M^2} \int d^8 \Omega\ \sum_{\xi ,\bar\xi }\ |J|^2\ ,
\eeq
$$
J=\int dx_0 d^d \vec x \ e^{i\omega x_0-i\vec k .\vec x} \ \xi^\mu \tilde \xi ^\nu\ T_ {\mu\nu}(x_0,\vec x)\ .
$$
For a classical string solution, the energy momentum tensor is
\beq
T_{\mu\nu} =\int d\sigma d\tau \ \delta^{(d)} (x-X(\tau,\sigma )) 
\p_- X^\mu \p_+ X^\nu \ .
\eeq
For the rotating ring (\ref{cring}),
\beq
J=\int d\sigma d\tau e^{2i N_0\tau -i \vec k .\vec X(\sigma,\tau ) } 
\xi^\mu\tilde \xi^\nu   \p_- X^\mu \p_+ X^\nu \ ,
\eeq 
$$
i\vec k.\vec X(\sigma ,\tau )=i \bar\k Z_1 +i\k  Z_1^* +i \bar {\td \k}Z_2 +i\td\k Z_2^*\ ,
$$
where $Z_1,\ Z_2$ are given in  (\ref{cring}).
This can be written as $J=J_R\ J_L$, with 
\beq
J_R=\int _0^{2\pi}d\sigma_- \ e^{iN_0\sigma_- - i\bar \k L e^{i\s_-}-i\k L e^{-i\s_-}} \big( i\bar \xe L e^{i\s_-} -i\xe L e^{-i\s_-}\big)\ ,
\eeq
\beq
J_L=\int_0^{2\pi} d\sigma_+ \ e^{iN_0\sigma_+-
i\bar {\td \k }L e^{i\s_+}-i\td \k L e^{-i\s_+}} 
\big( i\bar {\td \xe} L e^{i\s_+} -i \td \xe L e^{-i\s_+}\big)\ ,
\eeq
with $\s_\pm =\tau\pm \s $ and $L=\sqrt{2N}$.
Performing the integrals over $\s_+ $ and $\s_- $,we obtain
\beq
J_R= 2 N(-\sqrt{2N}i\k )^{N_0}
\sum_{s=0}^\infty {1\over s!} (-2\k \bar \k N)^s 
\bigg( {\bar \xe \k \over (N_0+s+1)! } +
{\xe \over 2N\k (N_0+s-1)!}\bigg)\ .
\eeq 
Remarkably, this coincides exactly with  (\ref{domi}) (up to an irrelevant phase $i^{N_0}$ which drops out in $|J|^2$).
It is important   to note is that the quantum calculation 
reduces to the classical
radiation process after taking the Stirling limit (\ref{stirg}), i.e. for large mass $M$
or large excitation number $N$. This is the limit where a classical description in terms of
a string soliton can be applied.

\subsection{Spectrum of massless emission} 



The spectrum of massless emission $M^5\times \Gamma (\omega )$
is shown in fig. 1  for different values $N=20,50,100$. 
A    salient feature of this spectrum is that the maximum is always at
$N_0=N-N'=3$ for all $N>9$.
The large $N$ form of the spectrum is given by (\ref{spw}), which in terms of $\omega $ reads
$$
\Gamma(\omega )=g_s^2M^{-5}  \bar f(2M\omega )\ .
$$
Thus the rotating ring decays mainly by emitting soft NS-NS massless
particles (gravitons and $B_{\mu\nu}$) of frequencies
$\omega=N_0/(2M)\sim M^{-1}$.

\begin{figure}[ht!]
\centering
\includegraphics*[width=250pt, height=180pt]{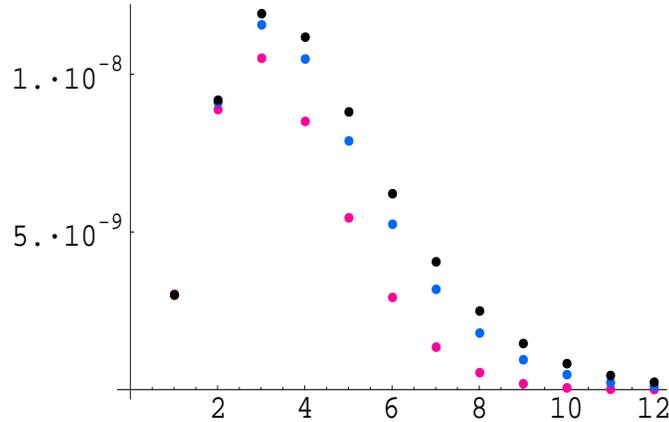}%
\caption{Spectrum  $M^5\times \Gamma (\omega )$
of massless emission from the rotating ring as a function of $N_0=2M\omega $,
 for  $N=M^2=20$ (lower points), $N=50$
and $N=100$ (higher points). Here $g_s^2=0.1$ and $\a'=4 $.}
\end{figure}

Since the dominant channel is the one where the final state
is a rotating ring of lower mass, the decay can be pictured
as a   rotating circular string
 whose radius gradually decreases as it loses energy by radiation.

For large $N_0$, the spectrum exponentially goes to zero. It is of the form
(see appendix C)
\beq
\Gamma(\omega )\cong N^{-{5\ov 2}}\ e^{-2N_0(2\log 2-1)}=N^{-{5\ov 2}}\ e^{-{\omega\over T_0}}\ ,\ \ \ 
\eeq
$$
T_0\equiv {a_0\over M}\ ,\ \ \ a_0^{-1}= 4(2\log 2-1)\ .
$$
The exponential form of the spectrum at large $\omega $, and the presence of a maximum
at finite frequency allows one to approximate the spectrum curve by
a thermal spectrum by introducing 
 a  ``grey body factor'' $\sigma_{\rm grey} (\omega )$
defined by
$$
\Gamma (\omega )={\rm const.}\ 
M^{-5}\omega ^{d-1} \sigma_{\rm grey}(\omega ,T_0 )\ 
{1\over e^{\omega\over T_0}-1}\ ,
$$
i.e. 
$$
 \sigma_{\rm grey}(\omega ,T_0)\equiv M^5 \Gamma (\omega )\omega^{1-d} 
(e^{\omega\over T_0}-1)\ .
$$
where $\Gamma(\omega )$ is the expression obtained in subsection 3.1.
The grey body factor here is introduced of course artificially, but given
the decay properties of the rotating ring, as a smooth radiation
process with frequencies peaked at $\omega\sim M^{-1}$,
 it is useful
to picture it as a thermal radiation process,
even if the exact spectrum formula is actually more complicated.

We have seen that the spectrum can be also obtained by a classical 
computation;
the fact that it is peaked at small frequencies is in agreement with the 
classical analysis of \cite{Damour} in the case of a smooth antenna.

%
%

We have also examined the spectrum of the string state with maximum angular
momentum $(b^\dagger \td b^\dagger )^N|0\rangle $. In this case we find that
the spectrum does not resemble a thermal spectrum, in particular,
it is not exponentially suppressed at large frequencies
(more precisely, in the region $1\ll N_0\ll N$).

\section{Presence of D branes}

\setcounter{equation}{0}

A question of interest is what is the lifetime of the 
rotating ring in the
presence of D branes. Here we show that  generically
the ring  cannot break into  open string fragments,  with Neumann or 
Dirichlet boundary conditions, nor in the presence of generic fluxes. 

Moreover the rotating ring cannot either decay into another
closed string plus open string fragments.
All these processes are exponentially suppressed, i.e they are of order 
$O(e^{-M^2})$.

There is an exceptional case in which the breaking  is not exponentially
suppressed, namely if there is a  D brane  tilted with an angle equal to $\pm {\pi / 4}$ with respect 
to the $Z_1$ and the $Z_2$ planes. This configuration is equivalent 
to a D brane such that its projection in the $Z_1, Z_2$ subspace 
appears as a D4 brane with a suitable flux.

\medskip

We begin by discussing the picture  by the classical world-sheet equations.
The classical solution corresponding to the circular string is given in 
eq.~(\ref{cring}), which we repeat for convenience:
\beq
X^{c}_j=G_j(\ta -\s )\ ,\ \ \ \ X^{c}_k=G_k(\ta +\s ) \ ,
\label{fring}
\eeq
where $j=1,2$ and  $k=3,4$, with $G_{1,3}(x )=L\ \cos(x) , \ 
G_{2,4}(x )=L\ \sin(x) $.

The classical equations are second order in $\tau$ therefore the breaking 
must
require continuity in $\tau$ of the classical functions and of
their first derivative (see \cite{IR2}).
In particular, this initial condition ensures that the Virasoro constraints
are satisfied for the outgoing fragments.
This initial condition holds for every coordinate. 

Take generically an open string to be of the form 
$X^{open}(\ta ,\s )=f(\ta +\s )+g(\ta -\s )$.
For  ${\cal N}$ Neumann or ${\cal D}$ Dirichlet boundary condition $g(x)=\pm f(x)$, but we consider
more general cases resulting from rotations or presence of fluxes. 

 Suppose that at  time $\tau =0$ the closed string breaks and
it becomes an open string.
The initial condition requires
\be
 Z^c(0,\s ) &= & Z^{open} (0,\s ) =f(\s )+g( -\s )\ ,
\label{conti}
\\ 
\p_{\tau } Z^c(0,\s ) &= &\mp\p_{\s } Z^c(0,\s )= 
\p_{\tau } Z^{open} (0,\s )= f'(\s )+g'(-\s ) \ .
\label{ffh}
\ee
with $\mp =- $ for the $j$ and $=+$ for the $k$ components. 
Since the continuity requirement (\ref{conti}) must hold for every 
$\s $, 
we can take the derivative with respect to $\s $ and get
\beq
\p_{\s } Z^c(0,\s ) =  f'(\s )-g'( -\s )\ .
\label{fgh}
\eeq
Combining (\ref{ffh}) with (\ref{fgh}) one finds $\{f'(x) \}_{m}=0$ for  $m=j$  and
$ \{ g'(x) \}_{m}=0$ for $m=k$ . 
Therefore the only possible final open strings are those of the form $g(\ta -\s )$ in the $Z_1$ plane
and $f(\ta +\s )$ in the $Z_2$ plane.

There is an exceptional case in which this is possible, namely  when in the $Z_1,Z_2$ subspace
there is a D brane that in the $Z_1,Z_2$ subspace it
appears as a D2 brane at an angle of $\pm {\pi / 4}$ 
with respect to the $Z_1,Z_2$ planes. 
Using prime coordinates to denote the rotated system which 
is aligned with the D brane (located at, say, $Z_{2'}=0$), one has
\beq 
Z_{1'}=\cos\theta\ g(\ta -\s )+\sin\theta\  f(\ta+\s )\ ,\ \ \ 
Z_{2'}=-\sin\theta\  g(\ta-\s )+\cos\theta\ f(\ta+\s )\ .
\eeq
The coordinate $Z_{1'}$ satisfies the  ${\cal N}$ boundary conditions
provided $ \cos\theta g(x)=\sin\theta  f(x)$.
The coordinate $Z_{2'}$ satisfies the  ${\cal D}$ boundary conditions
provided $\cos\theta  f(x)= \sin\theta g(x)=\sin\theta \tan\theta f(x)$, i.e.
provided $\tan^2\theta =1$, implying $\theta= \pm {\pi / 4}$.
Note that in the rotated system the ring solution becomes
\beq
Z_{1'}=\sqrt{2} L\ e^{i\tau}\cos\s\ ,\ \ \ \ 
 Z_{2'}=\sqrt{2} L\ e^{i\tau}\sin\s\ .
\eeq
Thus, with this  D brane configuration at a special angle, the circular
string can break into an open string.
Successively, the open string can easily break in more fragments.
 This configuration is equivalent to a D brane 
such that on the $Z_1,Z_2$ subspace it appears as a D4 brane 
with a  flux $F_{13}=F_{24}=1$.


\medskip

Quantum mechanically, a classically forbidden process can occur via tunnelling, 
but it will be exponentially suppressed for large mass (i.e. large $L$).
To see this, we use the boundary state formalism.
Omitting world-sheet fermions and ghosts, a general D brane is 
represented by the boundary state
\beq
|B\rangle=e^{-\alpha^{\dag}_{\cal N}\cdot\tilde\alpha^{\dag}_{\cal N}+
\alpha^{\dag}_{\cal D}\cdot\tilde\alpha^{\dag}_{\cal D}}|0\rangle\ ,
\eeq
where $\alpha_{\cal N,D}$ indicate the space coordinate mode operators with  
${\cal N}$ or ${\cal D}$ boundary conditions.
We restrict our attention to the $Z_1,Z_2$ subspace and consider  
a  D-brane such that its projection on it is 
a  D2 brane along the plane $X_{1'}X_{2'}$~:
\beq
|B\rangle =e^{-\alpha^{\dag}_{1'}\tilde \alpha^{\dag}_{1'}-\alpha^{\dag}_{2'}\tilde \alpha^{\dag}_{2'}+
\alpha^{\dag}_{3'}\tilde \alpha^{\dag}_{3'}+\alpha^{\dag}_{4'}\tilde \alpha^{\dag}_{4'}}|0\rangle\ .
\eeq
Here all $\alpha $ operators refer to the world-sheet mode operators of frequency 1;
the  coordinate indices $1',2', 3',4'$ appear as subindices for clarity in the notation. Write
\be
\alpha_{1'} &=& \cos(\theta )\alpha_{1}+\sin(\theta )\alpha_{3} \ ,\ \ \ \  
\alpha_{3'}=-\sin(\theta )\alpha_{1}+\cos(\theta )\alpha_{3} \ , \\ \nonumber
\alpha_{2'} &=& \cos(\theta )\alpha_{2}-\sin(\theta )\alpha_{4} \ ,   \   \  \  \  
\alpha_{4'} = \sin(\theta )\alpha_{2}+\cos(\theta )\alpha_{4}\ .
\ee
Then with our notation $b_{\pm}^\dagger =
{\alpha_{1}^\dagger \pm i\alpha_{2}^\dagger \ov\sqrt{2}}$ , 
$c_{\pm}^\dagger={\alpha_{3}^\dagger\pm i\alpha_{4}^\dagger\ov\sqrt{2}}$ 
we get, 
\beq
|B\rangle =e^{-\sin(2\theta )
(b_{+}^{\dag}\tilde c_{+}^{\dag}+c_{+}^{\dag}\tilde b_{+}^{\dag}+
             b_{-}^{\dag}\tilde c_{-}^{\dag}+c_{-}^{\dag}\tilde b_{-}^{\dag} )+
\cos(2\theta )(-b_{+}^{\dag}\tilde b_{-}^{\dag}-b_{-}^{\dag}\tilde b_{+}^{\dag}+
             c_{+}^{\dag}\tilde c_{-}^{\dag}+c_{-}^{\dag}\tilde c_{+}^{\dag} )}|0\rangle\ .
\eeq
Therefore, the amplitude for the process in which the ring is ``absorbed" by the brane 
thus becoming open strings on it, is
\beq
A=\langle 0|{(b_{+}\tilde c_{+})^N\ov N!}|B\rangle =(-\sin(2\theta ))^N\ .
\eeq
Thus this process is exponentially suppressed unless $\theta =\pm \pi /4$.

\medskip 

The rotating ring could still decay into open strings
in a two-step process in which 
the ring 
first decays into  a left-right symmetric (or antisymmetric) 
closed string state by emitting a closed string mode.
Then the resulting (anti) symmetric closed string state 
can decay into two open strings.

To calculate this process, we need to specify the emitted closed string state.
Emission of massive particles from the ring are exponentially suppressed. 
One could still wonder if the process could take place by emission of a massless particle. We now show that this process is even more suppressed, 
like $N^{-2N}$.

We first consider  an example, 
where we assume that the left-right symmetric
closed string state is 
$$
|\Phi'_{N'}\rangle \equiv {1\over (N'/2)!^2}
(b^\dagger c^\dagger )^{{N'\over 2}}(\tilde b^\dagger 
\tilde c^\dagger )^{{N'\over 2}}|0\rangle \ .
$$
Then we need to evaluate the matrix element
\beq
A_R={1\over (N'/2)! \sqrt{N!} }
\langle 0| (b c )^{{N'\over 2}} :\xi .\p X(1) e^{ik.X(1)}:
{b^\dagger }^N |0\rangle\ .
\eeq
The calculation is straightforward and the final result is
\beq
A_R =-i\sqrt{2}\xe \sqrt{N!} (-\sqrt{2}\kappa )^{N-{N'\over 2}-1 }(\sqrt{2}
\bar{\tilde \k} )^{N'\over 2}
 \sum_{r=0}^{N'/2} 
{(-2\k\bar\k )^r \over r!(N-{N'\over 2}+r-1)!({N'\over 2}-r)!}+\cdots
 \label{suma} 
\eeq
where dots stand for a similar  term which is of the same order.
To see the large $N$ behavior, one can consider the two cases, 
$N-N'=N_0$=fixed, or $N'$=fixed. Using the Stirling approximation, 
in both cases we find $A_R=O(N^{-N})$. 
One reason why this is so small 
is the factorial in the denominator  $(N-{N'\over 2}+r-1)!$ that starts
from $({N\over 2}+{N_0\over 2}-1)!$, which is large. 
The only way that this would not be large is if the final state is similar to the ring state insofar as it contains $b^{N_1}$ and $\tilde c^{N_2}$ 
with both  $N-N_{1}$ and $N-N_{2}$  small.
But such states are far from being left-right symmetric or antisymmetric. They are small deformations of the rotating ring.

One could consider more general emitted closed string states  of  the form $(b^\dagger \tilde b^\dagger )^{N_1}
(c^\dagger \tilde c^\dagger )^{N_2}|0\rangle $, $N_1+N_2=N'$. 
Since in this case we cannot have  both $N-N_{1}$ and $N-N_{2}$  small, we again expect
a similar suppression.
In particular, consider the extreme case
 $(b^\dagger \tilde b^\dagger )^{N'}$, i.e. $N_2=0$.
This represents a decay of the rotating ring into the closed string state 
of maximum angular momentum,
which, being a folded left-right symmetric string
may successively decay into open strings by breaking.
By a similar computation we find that the matrix element
 is of order $O(N^{-2N})$ for $N-N'=N_0$=fixed and of order $O(N^{-N})$
for $N'$=fixed.  
As expected for the reasons explained above, the process is suppressed.

\smallskip

Thus we conclude that, except for a  brane configuration at a special angle,
 the decay rate of the rotating ring 
in  presence of D branes is still dominated
by a decay into a similar ring state of lower mass by massless emission. 
The lifetime of the ring is therefore  given by the  formula of section 3.1, 
${\cal T}_{\rm ring} =c_0g_s^{-2} \ M^{5}$.

\section{Case of compactified dimensions } 

\setcounter{equation}{0}

The results can be easily extended to the case in which some $d_c$ dimensions have been compactified. 
When some dimensions are compactified, type II superstring theories can have
 many exactly stable
perturbative and non-perturbative states, associated with supersymmetric branes or supersymmetric fundamental strings wrapping cycles
(in ten uncompact dimensions, only the D0 branes of type IIA have finite mass).
Here we are interested in the stability of non-supersymmetric configurations. 
We will focus on the rotating ring, since this seems to be the most stable non-supersymmetric (zero charge) state.
We have explicitely considered the torus compactification, but the pattern is generic.
We assume that the radii $R_i$ of the compact dimensions are much smaller than the length $L=\sqrt{\alpha' N\ov 2}$ of our
circular string (its mass is $M={2\sqrt{N}\ov\sqrt{\alpha' }}$, $N$ being a large integer).
In the opposite case, the circular string does not feel the effect of the compactification and we have the same results as in the uncompactified case.

When two or more dimensions are compact, we have two different scenarios according to whether
the compact dimensions occur in both the $Z_1$ and $Z_2$ planes, where the circular string lies, or not.
In fact, in the former case, say that $ X_1$ in the $Z_1$ plane and $X_3$ in the $Z_2$ plane are compact, the circular string will wind 
around  $X_1$ with its Right part and around $X_3$ with its Left part, and it will easily 
break into winding modes. 

Instead, if the compact dimensions occur only in one of the  above planes, say in $Z_1$, then
 it is impossible for
the circular string to break into winding modes in the
$Z_1$ plane while respecting the Virasoro constraints, since classically the Left part in the $Z_2$ plane would remain unbroken. 
Quantum mechanically, this process is expected to be suppressed exponentially for large length of the circular string.

 This has been numerically verified by extending
the computation of \cite{CIR}
to the case of two compact dimensions \cite{CI}.

Therefore, if the compact dimensions occur only in one 
of the planes, say $Z_1$,  the dominant decay channel will be by
emission of radiation.

It is easy to extend our operatorial computation of Section 3 
to the case in which some $d_c$ dimensions are compactified. If no compact dimension occurs 
in the $Z_{1,2}$ planes, then the only modifications are
the integration over the angles, because now we have to integrate 
$\int d\Omega_{8-d_c}$ in the place of $\int d\Omega_{8}$, and in the phase-space 
factor, which is now $\omega^{7-d_c} M^{-2}\prod_i^{d_c} R_i^{-1}$ in the place of $\omega^{7} M^{-2}$, because the momentum of the 
emitted massless particle becomes $9-d_c$ dimensional. In fact we can neglect the possibility 
of emitting a particle with Kaluza-Klein momentum as it would be effectively massive and thus
exponentially suppressed.

The result is that the radiation spectrum is still described by a scaling expression
\beq
\Gamma \sim g_s^2{\alpha'}^{4}M^2\ \omega^{7-d_c}\left( \prod_i^{d_c} R_i^{-1}\right) \  f_{d_c}(\alpha' M\omega )\ .
\eeq
The form of $f_{d_c}(x)$ depends on $d_c$, the number of compactified dimensions.
The total radiation rate is obtained by summing over the integer $N_0={\alpha'\ov 2}M\omega $ getting
\beq
\Gamma_{\rm tot}\sim g_s^2 {\alpha'}^{(d_c-6)/2}M^{-(5-d_c)}\prod_i^{d_c}{ \sqrt{\alpha'}\ov R_i}\ .
\eeq

If some of the $d_c$ dimensions occur say in the $Z_{1}$ plane, the pattern is the
same as before, but the scaling function $f_{d_c}(x)$ is further modified by the fact
that some among the momentum components $k_1$ and $k_2$, which appear in
the expression giving  $f_{d_c}(x)$, are set to zero. If both the dimensions of the
$Z_{1}$ plane are compact and thus $k_1=k_2=0$, we see from the expression 
of the Right contribution to the spectrum (\ref{saa}), (\ref{saaa}) that  only $N_0=1$ survives, that is, 
the spectrum is in this case a single point:
\beq
\omega ={2\ov \alpha' M}\ .
\label{line}
\eeq
and the final state is  similar to the initial one with the change
$M^2={4\ov\alpha'}N \to M_1^2={4\ov\alpha'}(N-1)$.

Consider, in particular, $d_c=6$ (corresponding to $3+1$ uncompact dimensions).
In the case the circular string lies on the uncompact dimensions in its, say, Right part 
and on a compact plane in its Left part, and its length is much larger than every 
compactification radius, it decays  by emitting a massless particle of energy (\ref{line})
and changing its length from $\sqrt{\alpha' N\ov 2}$ to $\sqrt{\alpha' (N-1)\ov 2}$. 
This emission process takes a time
${\cal T}\sim {1\ov g_s^2M}\prod_i^{6}{ R_i\ov\sqrt{\alpha'}}$. 

The total time it takes to reduce its size from 
$N$ large to $N'$ of the order of some units, and thus easily breaking and disappearing, 
is obtained by integrating ${dN\over dt}\sim 1/{\cal T} $ from $N$ to $0$. We obtain
\beq
{\cal T}_{\rm tot}\sim N\ {\cal T}\cong {\alpha' M\ov g_s^2}\prod_i^{6}{ R_i\ov\sqrt{\alpha'}}\ .
\eeq

In a generic case in which some $6-d_{sc}$  radii are much larger than the circular string length $L$ and
$d_{sc}$ radii are 
of order $O(\alpha' )$, the total time it takes for the string to 
decay completely in massless modes is 
\beq
{\cal T}_{\rm tot}\sim {N\ov \langle N_0\rangle }\ {\cal T}\cong {1\ov g_s^2}{\alpha'}^{(4-{d_{sc}\over 2})}  M^{(7-d_{sc})}\prod_i^{d_{sc}}{ R_i\ov\sqrt{\alpha'}}\ .
\eeq
since now $N$ will decrease on average by a finite amount $\langle N_0\rangle$ of the order of unity in a single radiation step.

\section{Lifetime of average string state}
\setcounter{equation}{0}

The lifetime of the  average string state can be obtained
by computing the decay rate averaged over all initial states of the same mass $M$.
Here we will compute the average decay rate for 
massless emission, and neglect the decay channel into two massive particles.
Thus this will give an upper bound for the lifetime of the average string state.

The average decay rate for massless emission was computed in \cite{Amati2}
for the bosonic string. Here we  generalize this calculation to the type II superstring.

 We are going to compute an amplitude of the form
 \beq
\bar  \Gamma ={\rm const.}\ g_s^2{\omega^ 7\over M^2} 
\ \sigma_R\ \sigma_L\ ,\ \ \ \ 
\omega={M^2-{M'}^2\over 2M}={N_0\over 2\sqrt{N}}\ ,
\label{seiuno}
 \eeq
where
 \beq
 \sigma_R={1\ov \mathcal{N}_{N_R}}\oint_{C}{dz'\ov z'}z'^{-N'}\oint_{C'}{dz\ov z}
 z^{-N}{\rm Tr}[{\one+e^{i\pi F}\ov 2}V_R^\dag (k,1){\one+e^{i\pi F}\ov 2}z'^{\hat{N}}V_R(k,1)z^{\hat{N}}]\ .
\label{traza}
 \eeq
$V_R$ is the vertex operator (\ref{mvert}) and $\mathcal{N}_{N_R}$ is the number of
states at level $N=N_R=N_L$ in the NS-NS sector. Asymptotically,
$$
\mathcal{N}_{N_R}={\rm const.} \ N_R^{-{(D+1)\over 4}}
e^{a\sqrt{N_R}}\ ,
\ \ \ \ a=\pi\sqrt{D-2}\ ,
$$
with $D=10$ space-time dimensions.
We introduce a basis of coherent states:
 \beq
 |\{\lambda_n\}\rangle=\prod_{n=1}^\infty\exp[\lambda_n\cdot\alpha_n^\dagger]|0;p\rangle\ ,\ \ \ \ 
 |\{\theta _r\}\rangle=\prod_{r=1-\nu}^\infty\exp[\theta _r\cdot\psi_r^\dagger]|0;p\rangle\ ,
 \eeq
where $\theta_r$ are Grassmann variables. Defining
 \be
 \Sigma_{\bar\lambda,\lambda,\bar\theta ,\theta }& \equiv & {1\ov \mathcal{N}_N}\oint_{C}{dz'\ov z'}z'^{-N'}\oint_{C'}{dz\ov z}z^{-N}\\
 & \times & \langle \{\lambda_n\},\{\theta _r\}|{\one+e^{i\pi F}\ov 2}V_R^\dag (k,1){\one+e^{i\pi F}\ov
 2}z'^{\hat{N}}V_R(k,1)z^{\hat{N}}|\{\lambda_n\},\{\theta _r\}\rangle\ ,
 \nonumber
 \ee
 then the trace can be computed by
 integrating over all $\bar\lambda_n, \lambda_n$ and $\bar \theta_r, \ 
\theta_r $ with
 the weight factor $e^{-\bar\theta _r \cdot\theta _r}e^{-\bar\lambda_n\cdot\lambda_n}
$ due to normalization.

A shortcut to the calculation is to note that the following terms 
in (\ref{traza}) do not contribute to the trace:
\beq
\sum_{\Phi_N}\langle \Phi_N |
\xi.\psi k.\psi {z'}^{\hat N} \xi.\psi\ k.\psi |\Phi_N\rangle =0\ ,\ \ \ \ 
\sum_{\Phi_N}\langle \Phi_N | \xi.\psi \ k.\psi |\Phi_N\rangle =0\ .
\eeq
The reason why these terms vanish is that they must be proportional to
$\xi.k $ or $k^ 2$ by Lorentz invariance.
It is interesting to note that these correlators are not zero for particular
states such as the rotating ring, but the whole expression
 vanishes after taking the trace.
As a result, only the $\p X $ part of the vertex contributes.
Then, since there is no fermion insertion,  the fermion part in  
(\ref{traza})  gives 
the standard fermion partition function.
What remains is the bosonic part computed in \cite{Amati2} plus the ghost contributions, which are implicit in   (\ref{traza}).

In this case it is important to consider explicitely the ghosts, ensuring that we only keep physical states.
In the $bc$ ghost sector, we compute the amplitude (Right part)
 \beq
 \sigma_R=\sum_{\Omega_N, \Omega_{N'}}\langle \Omega_N|c(1)|\Omega_{N'}\rangle
 \langle\Omega_{N'}|c(1)|\Omega_N\rangle\ ,
 \eeq
where $c(1)$ is the ghost field associated with the vertex operator
$V(k,1)$ in (\ref{masslessspin2}) and $|\Omega_{N'}\rangle$ are
 ``physical'' ghost
 states (that is, such that the matter times ghost states satisfy the physical 
conditions).
 We then write:
 \beq
  \sum_{\Omega_{N'}} |\Omega_{N'} \rangle \langle\Omega_{N'}|=\oint {dz\ov z}
 z^{\hat N-{N'}}b_0(\sum_{\Psi} |\Psi \rangle
\langle\Psi|)=\oint {dz\over z}
 z^{\hat N-{N'}}b_0\one \ \ ,
 \eeq
where the $|\Psi\rangle$ are all states built with $b, c$ oscillators.
 The projector on states of level $N'$ is  the same of (\ref{Nlevelprojector})
(here we use its ghost part), the $b_0$ insertion is due to the fact
that while the sum $\sum_{\Psi}$ is over all of the ghost states, the
physical ones must satisfy the condition
 $b_0|\Omega\rangle=0$.
 These states, thanks to the $b,c$ algebra, can be written as
 $b_0|\Psi\rangle$ where $|\Psi\rangle$ is any ghost state.
Thus
we have  to compute
 \beq
 A={\rm Tr}[(-1)^{F_g}c(1)b_0c(z)z'^{\hat N}z^{\hat N}] \ ,
 \eeq
where $F_g$ anticommutes with all ghost modes.

This can be done with the same coherent state formalism
described above.
%
The final result is
 \beq
 A=\prod_{n=1}^\infty(1-w^n)^2,\quad w=z'z\ .
 \eeq

The superghost part of the average state computation gives directly the
superghost partition function (since the vertex operators $V(k,1),
V(k,z)$ do not carry any superghost contribution).


\smallskip

The result is thus \footnote{By the Jacobi identity,
one can further simplify this expression setting
${1\ov \sqrt{z}}g_3^8-{1\ov \sqrt{z}}g_4^8=g_2^8$.}
\be \label{finalaverage}
 \s_R  =  {\xi\cdot\xi (N-N')\over \mathcal{N}_N}\oint {dz \over z}
            {z^{-{N'}}f(z)^{2-D}\over 1-z^{N-N'}} 
      \left({1\ov
                       \sqrt{z}}g_3(z)^{D-2}-{1\ov
                       \sqrt{z}}g_4(z)^{D-2}+
                       g_2(z)^{D-2}\right) \ ,
\ee
where
 \beq
 f(z)=\prod_{n=1}^\infty (1-z^n)\,\ \ \
 g_3(z)=\prod_{r={1\ov 2}}^\infty(1+z^r)\ , \ \ \
 g_4(z)=\prod_{r={1\ov 2}}^\infty(1-z^r)\, \ \ \ 
 g_2(z)=\prod_{r=0}^\infty(1+z^r) \ .
\label{fss}
 \eeq
For large $N'$ the integral (\ref{finalaverage}) can be computed by a
saddle point approximation, the main contribution coming from $z\sim 1$. 
The computation is similar to the calculation of the asymptotic level
density $\mathcal{N}_N$ for level $N$.
 The behavior of the different functions in (\ref{fss})
 at  $z\sim 1$ follows from the modular properties of the theta functions. We find 
 \beq
 f(z)\cong {\rm const.} (1-z)^{-{1\ov 2}} e^{-{\pi^2\ov 6(1-z)}}\ ,\ \ \ \ 
 g_3(z)  \cong  {\rm const.}\ e^{{\pi^2\ov 12(1-z)}} \ , 
\eeq
\beq
 g_4(z)  \cong  {\rm const.}\ e^{-{\pi^2\ov 3(1-z)}}\ ,\ \ \  \ 
 g_2(z)  \cong {\rm const.}\ e^{{\pi^2\ov 12(1-z)}} \ .
 \eeq
 The saddle point is at
$
\ln z\cong -{a\ov 2\sqrt{N'}}
$
with $a=\pi\sqrt{D-2}$.
Using $N-N'=2\omega \sqrt{N}$, $\a'=4$, we obtain for large $N'$:
\beq
\s_R(\omega ) \cong \xi\cdot\xi\ \omega M \ {e^{a\sqrt{N-2\omega \sqrt{ N}}
-a\sqrt{N}}\over 1-e^{-a\omega }}\cong \xi\cdot\xi\ \omega M \ {e^{-a\omega}\over 1-e^{-a\omega }}\ .
\label{jju}
\eeq
Thus, by eq. (\ref{seiuno}), the average decay amplitude is given by 
\beq
\bar \Gamma ={\rm const.}\ g_s^2\xi^ 2 \td\xi^ 2 \ \s_{\rm grey}(\omega ,T) \ 
{1\over e^{\omega\over T}-1} \ \omega^8 \ ,
\eeq
where $T={1\over a\sqrt{\alpha' }}$ is the Hagedorn temperature of
type II superstring theory and $\s_{\rm grey} (\omega ) $ 
is a grey body factor given by
\beq
\s_{\rm grey}(\omega ,T)=
\omega {1-e^{-{\omega\over T}}\over  (1-e^{-{\omega\over 2T}})
(1-e^{-{\omega\over 2T}})}\ .
\eeq
Thus an average superstring state in type II superstring theory emits 
radiation as a grey body at the 
Hagedorn temperature. In the case of type I open superstrings, the spectrum 
for massless vector emission is obtained from the right part (\ref{jju}), i.e.
\beq
\bar \Gamma_{\rm  I}(\omega ) \cong \xi\cdot\xi\  M^{-1} \ 
{e^{-{\omega\over T_{\rm I}}}\over 
1-e^{-{\omega\over T_{\rm I} }} }\ \omega^8\ ,
\label{jjuu}
\eeq
where $T_{\rm I}={2\over a\sqrt{\alpha' }}$ is the Hagedorn temperature of
type I superstring theory. Hence average states in type I superstrings
emit radiation as a perfect black body. 

\smallskip

To compute the total average decay rate, one sums over all $N_0$. 
Recalling that $\omega=N_0/2M$, this sum can be approximated by an integral
over $\omega $ by introducing the differential $2Md\omega $.
We find
\beq
\bar \Gamma_{\rm tot} \sim g_s^ 2\ M\ .
\eeq
Since $\bar\Gamma $ represents the average decay rate taking into account massless emission only,  
the  lifetime of an average state has the upper bound
\beq
\bar {\cal T} \leq  g_s^{-2}\ M^{-1}\ .
\eeq
The same result holds for the bosonic and heterotic string theory.\footnote{
Since in open string theory the decay in massive states is important, because
breaking is never suppressed, the bound put by
the massless emission result $\bar {\cal T}_{\rm type\ I} \leq
{\rm const. } \sqrt{\a' }$
in this case could be rather weak.}

\section{Discussion}

The average lifetime computed here,
$\bar {\cal T}\leq M^{-1}$, shows that most of the massive states 
at each mass level have a 
 short lifetime, which decreases for larger masses.
There can only be a few states at each mass level whose 
lifetime increases with the mass.
Some of these few states are given by 
the family of states $|\Phi_{n,k}\rangle $, $N=2k$, studied in
\cite{CIR}, interpolating between the rotating ring $n=k$ and the state $n=0$.
The corresponding classical configurations describe rotating elliptical strings, which become
circular for $n=k$ and a folded, straight closed string for $n=0$.
For generic $n$ one finds that the lifetime
${\cal T}$ also increases with the mass, though at 
much lower rate for states that significantly differ
from the rotating ring (i.e. states with  $n$ not close to $k$).
When $n \ll k$, the ellipse is very eccentrical and the decay by breaking into two massive string states
becomes the dominant channel.
These are all states of high angular momentum.

The law ${\cal T}_{\rm ring}\cong {\rm const.}\ M^5 $ has been derived in four independent ways.
1) By evaluation of the imaginary part of the two-point function \cite{CIR},
2) by operator calculation of the full massless emission rate, 3) by operator calculation of the dominant
decay channel into a ring of lower mass and 4) by classical radiation from the rotating string soliton configuration.
The latter also confirms the identification of the quantum state with the string soliton.

The rotating ring has the special feature that it decays predominantly into another rotating ring, i.e.
the decay can be viewed as a rotating circular string whose radius gradually 
decreases as it emits radiation.
The decay rate of the rotating ring has the scaling 
property $\Gamma =c g_s^2\ M^{-(5-d_c)}\ \bar f(N_0)$, where $N_0=M^2-{M'}^2=2\omega M$,
$d_c=$ number of compact dimensions with radius $R<<\alpha' M$,
and a spectrum sharing some qualitative features with a thermal spectrum, such as an exponential tail.
This is not the case e.g. for the
 state of maximum angular momentum $(b^\dagger \td b^\dagger )^N|0\rangle $. 

In the case of the average state, one finds as in \cite{Amati2} that the spectrum becomes exactly thermal
in the large mass limit 
and that the average type II superstring emits radiation as a grey body with temperature
equal to the Hagedorn temperature (while
the average type I superstring  behaves as a perfect black body at the Hagedorn temperature).

\section*{Acknowledgements}

We would like to thank J.L. Ma\~ nes 
for pointing out a missing factor $1/M^2$ in
(6.1)  in a previous version of this work. This factor follows from
 eq.~(\ref{dcy}) and corrects the average lifetime by a factor $M^2$.
We acknowledge partial support by
the European Community's Human potential
Programme under the contract HPRN-CT-2000-00131.
The work of J.R. is
supported in part also by MCYT FPA,
2001-3598 and CIRIT GC 2001SGR-00065.
J.R. would like to thank SISSA and CERN for hospitality during the course of this work.



\setcounter{section}{0}
\appendix{Notation}

We consider type II superstring theory in the NS formulation.
The world-sheet action is given by
\be
S={1\over 2\pi \alpha '}\int d^ 2z\big( \partial X^\mu \bar \partial X_\mu + \psi ^\mu
\bar\partial \psi_\mu + \tilde \psi^\mu \partial \tilde \psi_\mu \big)
\ee
We will be interested in string states corresponding to excitations 
in two planes,
$$
Z^1=\frac{X^1+i X^2}{\sqrt{2}}, \ \ Z^2=\frac{X^3+i X^4}{\sqrt{2}}\ ,
$$ 
$$
\psi^{Z_1}=\frac{\psi^1+i \psi^2}{\sqrt{2}}, \ \
\psi^{Z_2}=\frac{\psi^3+i \psi^4}{\sqrt{2}}\ .
$$
The expansion for the bosonic world sheet fields are as follows
$$
Z_R^1 = {i} \sqrt{\frac{\a'}{2}}\sum_{n=1}^\infty {1\over \sqrt{n}}
\big[ b_{n-} z^{-n} - b_{n+}^\dagger  z^{n}\big]\ ,\ \ \ 
Z_L^1 = {i} \sqrt{\frac{\a'}{2}}\sum_{n=1}^\infty {1\over \sqrt{n}}
\big[ \tilde b_{n-} \bar z^{-n} - 
\tilde b_{n+}^\dagger  \bar z^{n}\big]\ ,
$$
$$
Z_R^2 = {i} \sqrt{\frac{\a'}{2}}\sum_{n=1}^\infty {1\over \sqrt{n}}
\big[ c_{n-} z^{-n}  - c_{n+}^\dagger z^{n} \big]\ ,\ \ \ 
Z_L^2 = {i} \sqrt{\frac{\a'}{2}}\sum_{n=1}^\infty {1\over \sqrt{n}}
\big[ \tilde c_{n-} \bar z^{-n}- 
\tilde c_{n+}^\dagger \bar z^{n} \big]\ ,
$$
$$
[b_{n\pm},b_{m\pm }^\dagger]=\delta_{nm}\ ,\ \  
[\tilde b_{n\pm},\tilde b_{m\pm }^\dagger]=\delta_{nm}\ ,\ \
[c_{n\pm},c_{m\pm }^\dagger]=\delta_{nm}\ ,\ \  
[\tilde c_{n\pm},\tilde c_{m\pm }^\dagger]=\delta_{nm}\ .
$$
For the fermion fields, we have
\be
\psi ^{Z_i}=  \sqrt{\alpha'\over 2} \sum_r \psi_r^{Z_i}  z^{-r}\ ,
\ \ \ \td \psi ^{Z_i}=  \sqrt{\alpha'\over 2} \sum_r \td \psi_r^{Z_i}  \bar 
z^{-r}\ ,
\ee
$$
\{ \psi_r^{Z_i}, \psi_s^{Z_j}\} =\delta_{rs} \delta_{ij}\ .
$$
The mass formula is given by
\beq
\a' M^2= 2(\hat N_R+\hat N_L)\ ,\ \ \ \ \  N_R=N_L \ ,
\label{masa}
\eeq
where the number operators are
\beq
\hat N_R=\sum_{n=1}^\infty n( b_{n+}^\dagger b_{n+}+ b_{n-}^\dagger b_{n-}
+ c_{n+}^\dagger c_{n+}+ c_{n-}^\dagger c_{n-})+\sum_{r={1\over 2}}^ \infty
r\psi_{-r}^{Z_i} \psi_r^{Z_i}- a_0\ . 
\eeq
The expression for $\hat N_L$ is similar, with the obvious changes.
The normal ordering constant $a_0$ is $a_0=1/2$ in the NS sector and $a_0=0$ in the R sector.

The basic correlators are
\beq
\langle \partial_w  X(w)\partial_z X(z)\rangle =
- {\alpha'\over 2}{zw\over (w-z)^ 2} \ ,\ \ \ \ \ 
\langle  \psi(w)\psi(z)\rangle ={\alpha'\over 2}{ (zw)^{1/2}\over w-z} \ .
\eeq

We use the following angular coordinates
$$
\hat x_1=\sin\theta \cos\alpha \cos\beta\cos\varphi\ ,\ \ \ \ 
\hat  x_2=\sin\theta \cos\alpha \cos\beta\sin\varphi\ ,
$$
$$
\hat x_3=\sin\theta \cos\alpha \sin\beta\cos\psi\ ,\ \ \ \ 
\hat x_4=\sin\theta \cos\alpha \sin\beta\sin\psi\ ,
$$
$$
\hat x_5=\sin\theta \sin\alpha \cos\beta'\cos\varphi'\ ,\ \ \ \ 
\hat x_6=\sin\theta \sin\alpha \cos\beta'\sin\varphi'\ ,
$$
$$
\hat x_7=\sin\theta \sin\alpha \sin\beta'\cos\psi'\ ,\ \  
\hat x_8=\sin\theta \sin\alpha \sin\beta'\sin\psi'\ ,\ \ \hat x_9=\cos\theta \ .
$$
Thus the volume element is 
$$
d^8\Omega =\sin^7 \theta \cos^ 3\alpha \sin^3\alpha \cos\beta \sin\beta 
\cos\beta '\sin\beta' 
d\theta d\alpha d\beta d\varphi d\psi  d\beta' d\varphi' d\psi'  
$$


In order to sum over polarizations, we use the  relations
\be
\sum_\xi \xi^i \xi^j =\delta^ {ij}-{k^i k^j\over |\vec k|^ 2}\ ,\ \ \ \ 
\sum_\xi \xi^i \xi_i =d-1=8\ .
\ee
This gives ($\k $ and $\xe $ are defined in (\ref{cpx}), (\ref{cpxx})~)
\be 
\sum_\xi \xe^ 2= -{\k^ 2\over  |\vec k|^ 2}\ ,\ \ \ \ 
\sum_\xi \bar \xe^ 2 = -{\bar \k^ 2\over  |\vec k|^ 2}\ ,\ \ \ 
\sum_\xi \xe \bar \xe =1-{\k\bar \k\over  |\vec k|^ 2}\ .
\ee
Writing
$$
2\k^ 2={N_0^ 2\over 4 N}\ e^{2i\varphi}v^ 2\ ,\ \ \ v^ 2\equiv
\sin^ 2\theta\cos^ 2\alpha\cos^ 2\beta\ ,
$$
we have
\be 
\sum_\xi \xe^ 2 = - e^{2i\varphi} {v^ 2\over  2}\ ,\ \ \ \ 
\sum_\xi \bar \xe^ 2 = -\ e^{-2i\varphi} {v^ 2\over  2},\ \ \ 
\sum_\xi \xe  \bar \xe =1-{v^ 2\over   2}\ .
\ee

\appendix{Other contributions to the decay rate of the rotating ring}
\setcounter{equation}{0}
The remaining contributions  to the decay rate of the rotating ring
indicated in section 3.1 
 are
\be
\sigma_2 &=& -i{1\over N!} 
 \oint {dz\over z}\ z^{N_0} 
\langle 0|b^N e^ {-ik.X(1)} e^{ik.X(z)}\xi .\p X(1)
{b^\dagger}^N | 0 \rangle 
\langle 0|\psi_{1\over 2}^{\bar Z} \xi .\psi(z)  k.\psi(z) \psi_{-{1\over 2}}^Z | 0 \rangle \ ,
\nonumber\\
\sigma_3 &=& i{1\over N!} 
 \oint {dz\over z}\ z^{N_0} 
\langle 0|b^N e^ {-ik.X(1)} e^{ik.X(z)}\xi .\p X(z)
{b^\dagger}^N | 0 \rangle 
\langle 0|\psi_{1\over 2}^{\bar Z} \xi .\psi(1)  k.\psi(1) \psi_{-{1\over 2}}^Z | 0 \rangle\ ,
\nonumber\\
\sigma_4 &=& - {1\over N!} 
 \oint {dz\over z}\ z^{N_0} 
\langle 0|b^N e^ {-ik.X(1)} e^{ik.X(z)}
{b^\dagger}^N | 0 \rangle 
\langle 0|\psi_{1\over 2 }^{\bar Z} \xi .\psi(1) 
 k.\psi(1)\ \xi .\psi(z)  k.\psi(z) \psi_{-{1\over 2}}^Z | 0 \rangle \ ,
\nonumber\\
\sigma_5 &=& {1\over N!} 
 \oint {dz\over z}\ z^{N_0 } 
\langle 0|b^N e^ {-ik.X(1)} e^{ik.X(z)}
{b^\dagger}^N | 0 \rangle 
\langle 0|\xi .\p X(1)\xi . \p X(z) | 0 \rangle
 \langle 0|\psi_{1\over 2}^{\bar Z}\psi_{-{1\over 2}}^Z | 0 \rangle 
\nonumber
\ee
Computing the matrix elements, we find
\be
\sigma_2= -4
 \oint {dz\over z}z^{N_0}\big( (\bar\xe\xe\bar \k \k-\bar \xe^2  \k ^ 2) z^{-1}
+(\xe^2\bar \k^2-\bar \xe \xe
\bar  \k \k) \big)
\sum_{m=0}^ {N-1}
{N!(2\k\bar \k )^ m \over m!(m+1)!(N-m-1)!}{(1-z)^ {2m+1}\over z^ m}\ ,
\non
\ee
\be
\sigma_4=4 \xi \cdot \xi \ \k\bar \k
 \oint {dz\over z}z^{N_0}\sum_{m=0}^ {N}
{N!(2\k\bar \k )^ m \over m!^ 2(N-m)!}\big[{(1-z)^ {2m-1}\over z^ m}-{(1-z)^ {2m-1}\over z^ {m-1}}\big]\ ,
\non
\ee
\be
\sigma_5=2  \xi \cdot  \xi
 \oint {dz}z^{N_0}\sum_{m=0}^ {N}
{N!(2\k\bar \k )^ m \over m!^ 2(N-m)!}{(1-z)^ {2m-2}\over z^ m}\ ,
\non
\ee
and $\s_3=\s_2^*$.
Expanding the binomials $(1-z)$ and computing the integral over $z$, we find
\be
\sigma_2= 4\!\!
\sum_{m=N_0-1}^ {N-1}\!
{(-1)^{m-N_0}
N!(2m+1)!(2\k\bar \k )^ m \big[ 
(\bar \xe\xe \bar  \k \k\!\!-\!\!\xe^2\bar \k^2) (m-\!\!N_0+\!\!1)\!\!
-\!\!(\bar\xe^2 \k^2\!\!-\!\! \bar \xe\xe\bar \k \k )(m+N_0+1)
\big]  \over m!(m+1)!(N-m-1)!(m-N_0+1)!(m+N_0+1)!}\ ,
\non 
\ee
\be
\sigma_4=4 \xi \cdot \xi\ \k\bar \k
\sum_{m=N_0}^ {N}
{(-1) ^{m-N_0}  N!(2m)!(2\k\bar \k )^ m \over m!^ 2(N-m)! (m-N_0)!(m+N_0)!  }\ ,
\non
\ee
\be
\sigma_5=-2  \xi \cdot  \xi
\sum_{m=N_0+1}^ {N}
{ (-1)^{m-N_0} N!(2m-2)!\ (2\k\bar \k )^ m \over m!^ 2(N-m)!(m+N_0-1)!(m-N_0-1)!}\ .
\non
\ee
Now we can estimate which is the dominant contribution in the large $N$ limit,
with fixed $N-N'$.
Using Stirling formula (\ref{stirg}) 
and the fact that $\k\k \sim N^{-1}$, we see that the dominant term is 
$\sigma_1 $, with $\sigma_1=O(N)$. Similary, we see that  
$\sigma_2,  \sigma_3, \sigma_5  $ are of order $O(1)$ and $ \sigma_4$ is of order $O(N^{-1})$
(the explicit numerical evaluation shows that $\s _2, ..., \s_5$ are
two or more orders of magnitudes  smaller than $\s_1 $ already for $N=20$).
The application of the Stirling formula is justified in the present case 
because decays with $N_0\gg 1$ are always suppressed as long as $N$ is large.

\appendix{Tail of the spectrum } 
\setcounter{equation}{0}

We consider the expression for $\Sigma_{11}$ (\ref{nombre}).
Using the Stirling formula (\ref{stirg}), we find
\be
\Sigma_{11}= \sqrt{\pi } N^ 2 x^{2N_0}
\sum_{r,s=0}^\infty 
{   (-x)^{r+s}\over \Gamma(2N_0+r+s+{13\over 2})}
a(r)a(s)\ ,
\ee
$$
a(r)={(2r+2N_0)!\over r! (N_0-1+r)!(r+2N_0)!}\ ,
$$
where $x\equiv 2\k\bar \k N={N_0^ 2\over 4}$.
We are interested in the large $N_0$ behavior of this sum.
Because the series converges very rapidly,
we can approximate
\be
{(2N_0+2r)!\over (2N_0+r)!}\cong (2N_0)^ r\ ,\ \ \ 
{(2N_0+2s)!\over (2N_0+s)!}\cong (2N_0)^ s\ .
\ee
At the same time, we use Stirling formula to approximate
\be
 \Gamma(2N_0+r+s+{13\over 2}) &\cong&
 (2N_0)! \ (2N_0)^ {r+s+{11\over 2}} \nonumber\\
&\cong &
\sqrt{2 \pi } (2N_0)^{2N_0+r+s+6}\ e^{-2N_0}\ .
\ee
We find
\be
\Sigma_{11}\cong N^2 {x^{2N_0}\over \sqrt{2} }{ e^{2N_0}\over (2N_0)^{2N_0+6}}
\ S(N_0)^2\ ,
\ee
\be
S(N_0)\equiv \sum_{r=0}^\infty {(-x)^r\over r!(N_0-1+r)!}\ .
\ee
We can write this sum as
\be
S(N_0)={1\over 2\pi i}\oint {dt\over t} \ t^{N_0-1}\ \exp [-x t+{1\over t}]\ ,\ \ \ \ x={N_0^2\over 4}\ .
\ee
The integral can be computed by saddle-point approximation, 
with the result
\be
S(N_0)={\rm const.}\ N_0^{\beta}\ e^{N_0\log2-N_0\log N_0}\ ,
\ee
where $\beta $ is a constant of order one.
Adding the phase space factor,  we finally obtain
\be
\Gamma_{11}={1\over M^2} \omega ^ 7 \Sigma_{11}=
  N \big({N_0\over 2\sqrt{N}}\big) ^7
N_0^ {2\beta-6}  e^{-2N_0(2\log 2-1)}\ ,
\ee
i.e.
\be
\Gamma_{11}=N^{-{5\over 2}}\ N_0^{1+2\beta}\ e^{-2N_0(2\log 2-1)}\ .
\ee
Similarly, one can consider $\Sigma_{22}$
and $\Sigma_{12}$. We find the same behavior.

\vskip 2cm




 \end{document}